# The spatial meaning of Pareto's scaling exponent of city-size distribution

Yanguang Chen

Department of Geography, Peking University, Beijing 100871, PRC. E-mail: chenyg@pku.edu.cn

**Abstract**: The scaling exponent of a hierarchy of cities used to be regarded as a fractal parameter. The Pareto exponent was treated as the fractal dimension of size distribution of cities, while the Zipf exponent was treated as the reciprocal of the fractal dimension. However, this viewpoint is not exact. In this paper, I will present a new interpretation of the scaling exponent of rank-size distributions. The ideas from fractal measure relation and the principle of dimension consistency are employed to explore the essence of Pareto's and Zipf's scaling exponents. The Pareto exponent proved to be a ratio of the fractal dimension of a network of cities to the average dimension of city population. Accordingly, the Zipf exponent is the reciprocal of this dimension ratio. On a digital map, the Pareto exponent can be defined by the scaling relation between a map scale and the corresponding number of cities based on this scale. The cities of the United States of America in 1900, 1940, 1960, and 1980 and Indian cities in 1981, 1991, and 2001 are utilized to illustrate the geographical spatial meaning of Pareto's exponent. The results suggest that the Pareto exponent of city-size distribution is not a fractal dimension, but a ratio of the urban network dimension to the city population dimension. This conclusion is revealing for scientists to understand Zipf's law and fractal structure of hierarchy of cities.

## 1. Introduction

The rank-size distribution of cities in a large geographical region always follows Zipf's law, unless the sphere of influence of the largest city is far greater than area of the region. Zipf's law for cities is one of the most conspicuous empirical facts in the social sciences (Gabaix, 1999). It is easy to demonstrate that Zipf's law is mathematically equivalent to Pareto's law (Chen, 2012a). The rank-size rule suggests a kind of fractals, and Zipf's exponent used to be regarded as the reciprocal of a similarity dimension (Mandelbrot, 1982). In theory, the Zipf scaling exponent is just the reciprocal of the Pareto scaling exponent; therefore, the Pareto exponent was regarded as the fractal dimension of rank-size distributions. However, this viewpoint is inexplainable for cities. The problem is that the size measure of cities (e.g. city population, urban area) is not a linear scale defined in a 1-dimensional Euclidean space. The necessary condition of defining a fractal dimension using scaling relation between two measurements (e.g. length, area, number) is that one of the measurements represents a 1-dimensional scale.

In order to calculate a fractal dimension of a system, we must make use of a geometrical measure relation. If and only if the relation between one measurement and another measurement of the system follows the scaling law, the scaling exponent (a power of a measure) can be treated as a fractal parameter; if and only if one of the two measures is a linear scale defined in a 1-dimensional space, the power of the linear measure can be considered to be the fractal dimension of another measure. Otherwise, the power exponent is a ratio of one fractal dimension to another fractal dimension rather the fractal dimension itself. If the Pareto exponent of city-size distribution was a fractal dimension, two questions would arise: whether or not the city size is a 1-dimension measure which can be treated as a linear scale? How to interpret the spatial meaning of the fractal dimension of city-size distribution? If and only if the two questions are replied, the Pareto exponent as well as the Zipf exponent can be made clear in urban studies.

The above two question cannot be answered in light of the traditional concepts of cities. The precondition that the Pareto exponent can be regarded as a fractal dimension is that city size is a 1-dimensional measure. Urban population is always employed to measure city size, but the population size of a city is not 1-dimensional measure (Lee, 1989; Nordbeck, 1971). Based on the digital map defined in a 2-dimensional Euclidean space, the dimension of city population comes between 1 and 2 (Chen, 2008). This gives rise to another problem. Empirically, the Pareto exponent of the population size distribution of cities is always close to unit. The rank-size distribution is mathematically equivalent to a self-similar hierarchy (Chen, 2012b), and a hierarchy and a network represent two different sides of the same coin (Batty and Longley, 1994).

This suggests that the fractal dimension of city number based on given population size is just the dimension of a city network in a 2-dimensional geographical space. In this case, the fractal dimension of city-size distribution should fall into 1 and 2 rather than approach 1. However, the empirically observed values of the Pareto exponent are always near 1 instead of coming between 1 and 2.

In this paper, the Pareto exponent of city-size distribution will be reinterpreted using the ideas from fractals and the principle of dimension consistency. I will demonstrate that the Pareto exponent is a ratio of the fractal dimension of a network of cities to the average dimension of city population in the network. Accordingly, the Zipf exponent can be readily understood since Zipf's law is theoretically the inverse function of Pareto's law. The rest parts of this work are organized as follows. In Section 2, the Pareto exponent of city-size distribution will be demonstrated to be fractal dimension ratio rather than a similarity dimension itself. In Section 3, the geographical spatial implication behind the Pareto exponent will be revealed, and two case studies will be provided to help readers understand the Pareto exponent. In Section 4, several related questions will be discussed so that the geometrical meaning of the Pareto exponent becomes clearer. The paper will be concluded by summarizing the main points of this study. Because of the equivalence relation between Zipf's law and Pareto's law and the understandability of Zipf's law, the mathematical description will start from Zipf's distribution of cities.

## 2. Fractal dimension of city-size distribution

### 2.1 Rank-size distribution and Zipf's law

If the cities within a geographical region comply with the rank-size rule, the size distribution of cities can be described with Zipf's law as below

$$S(k) = S_1 k^{-q}, \qquad (1)$$

where $k$ refers to the rank of cities in a descending order, $S(k)$ to the size of the city of rank $k$, $q$ denotes the Zipf scaling exponent, and the proportionality coefficient $S_1$ is the size of the largest city in theory. The size can be measured with city population, urban area, and so on. The inverse function of equation (1) is $k=(S(k)/S_1)^{-1/q}$, in which the rank $k$ represents the number of the cities with size greater than or equal to $S(k)$. Reducing $S(k)$ to $S$ and substituting $k$ with $N(S)$, we have

$$N(S) = \eta S^{-p}, \tag{2}$$

which is equivalent to Pareto's law. In equation (2), the power $p=1/q$ denotes the Pareto scaling exponent, and $\eta=S_1^{1/q}$ is the proportionality coefficient.

The Pareto scaling exponent $p$ is always regarded as the fractal dimension of city-size distribution, which is defined in 1-dimension space (Chen, 2012a;Chen, 2012b; Frankhauser, 1990; Mandelbrot, 1982; Nicolis *et al*, 1989). However, two problems arise. First, the scaling exponent $p$ is not a real fractal dimension because the size measurement $S$ is not a basic scale. If and only if the Euclidean dimension of a measurement is $d=1$, the measurement can act as the basic scale to define a fractal dimension. The examples of basic scale are as follows: the sidelength of a square, the radius of a circle, the span of a divider, the length of a yardmeasure…. Second, the spatial meaning of the scaling exponent is not clear. In other words, if the Pareto scaling exponent is a type of fractal dimension, how to understand it from the viewpoint of geographical space? In order to reveal the spatial implication of the fractal dimension of city-size distribution, it is necessary to draw an analogy between hierarchies of cities and regular fractal hierarchies (Figures 1 and 2).

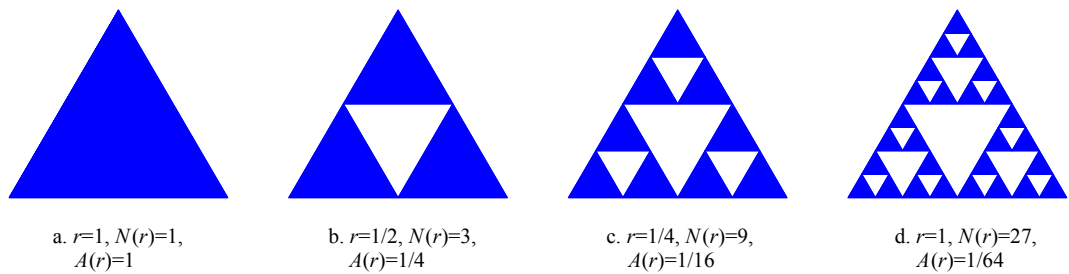

a. $r=1$, $N(r)=1$, $A(r)=1$    b. $r=1/2$, $N(r)=3$, $A(r)=1/4$    c. $r=1/4$, $N(r)=9$, $A(r)=1/16$    d. $r=1$, $N(r)=27$, $A(r)=1/64$

**Figure 1 The first four steps of the Sierpinski gasket**

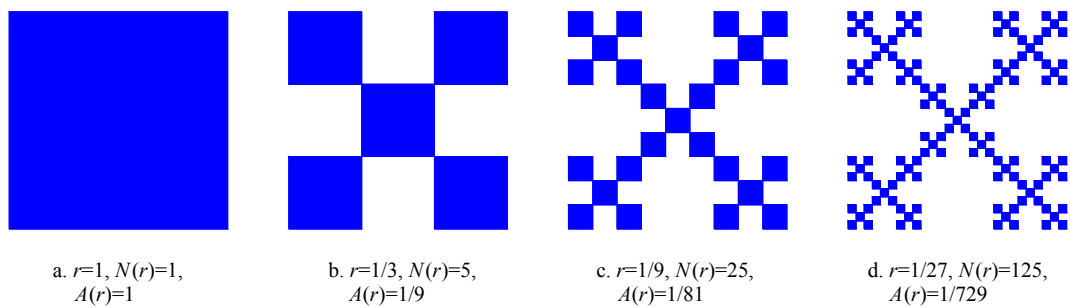

a. $r=1$, $N(r)=1$, $A(r)=1$    b. $r=1/3$, $N(r)=5$, $A(r)=1/9$    c. $r=1/9$, $N(r)=25$, $A(r)=1/81$    d. $r=1/27$, $N(r)=125$, $A(r)=1/729$

**Figure 2 The first four steps of the Jullien-Botet growing fractal**

## 2.2 Regular fractal hierarchies

A fractal is a hierarchy with cascade structure, which is similar to the hierarchy of cities. The similarity dimension of a regular fractal can be given by the following formula:

$$N(r_m) = N_0 r_m^{-D}, \qquad (3)$$

where $m$ denotes the level of a fractal hierarchy and corresponds to the step of fractal generation ($m$=0,1,2,…), $r_m$ refers to the linear size of the fractal copies at the $m$th level, $N(r_m)$ to the number of fractal copies with a linear scale of $r_m$, $N_0$ is the proportionality coefficient, and $D$ is the similarity dimension. Generally speaking, we have $N_0$=1. For many regular fractals, the similarity dimension equals it box dimension. If the area of a fractal copy at the $m$th level is notated as $A_m$, we will have $A_m \propto r_m^2$, where $\propto$ indicates "be directly proportional to". If the symbol $N(r_m)$ is reduced to $N_m$, equation (3) will be rewritten as

$$N_m = \mu A_m^{-D/2}, \qquad (4)$$

where the proportionality constant $\mu = N_1 A_1^{D/2}$. Generally, we have $\mu$=1. Equation (4) suggests a geometrical measurement relation as follows

$$N_m^{1/D} \propto A_m^{-1/2}, \qquad (5)$$

which is in fact a inverse allometric scaling relation (Chen, 2010).

Two simple regular fractals can be employed to illustrate equations (3), (4) and (5): one is Sierpinski gasket displayed in Figure 1, and the other, the Jullien-Botet growing fractal displayed in Figure 2 (Jullien and Botet, 1987; Vicsek, 1989). Partial data of the two fractal hierarchies are tabulated as below (Table 1). From equation (3) it follows a fractal dimension formula such as

$$D = \frac{\ln(N_{m+1}/N_m)}{\ln(r_m/r_{m+1})}. \qquad (6)$$

For the Sierpinski gasket, the fractal dimension is $D=\ln(3)/\ln(2)\approx1.585$ in terms of equation (6). This value can be derived from the inverse allometric function. By equation (4) or (5), the geometrical measurement relationship between the area of fractal copies $A_m$ and the number of fractal copies $N_m$ at the $m$th level is

$$N_m \propto A_m^{-D/2} = A_m^{-0.7925}.$$

So the fractal dimension is $D$=2*0.7925=1.585. For the Jullien-Botet growing fractal, the similarity dimension is $D=\ln(5)/\ln(3)\approx 1.465$ according to equation (6). By equation (4) or (5), the geometrical measurement relationship between $A_m$ and $N_m$ is

$$N_m \propto A_m^{-D/2} = A_m^{-0.7325}.$$

Thus the fractal dimension is $D=2*0.7325=1.465$. The fractal measurement relationships can be illustrated with log-log plots based on the first ten steps (Figure 3).

**Table 1 The linear scale, area, and number of fractal copies in each levels of two regular fractal hierarchies (the first 10 levels)**

| Step | Sierpinski gasket | | | Jullien-Botet growing fractal | | |
|---|---|---|---|---|---|---|
| $m$ | $r_m$ | $A_m$ | $N_m$ | $r_m$ | $A_m$ | $N_m$ |
| 0 | 1/1 | 1/1 | 1 | 1/1 | 1/1 | 1 |
| 1 | 1/2 | 1/4 | 3 | 1/3 | 1/9 | 5 |
| 2 | 1/4 | 1/16 | 9 | 1/9 | 1/81 | 25 |
| 3 | 1/8 | 1/64 | 27 | 1/27 | 1/729 | 125 |
| 4 | 1/16 | 1/256 | 81 | 1/81 | 1/6561 | 625 |
| 5 | 1/32 | 1/1024 | 243 | 1/243 | 1/59049 | 3125 |
| 6 | 1/64 | 1/4096 | 729 | 1/729 | 1/531441 | 15625 |
| 7 | 1/128 | 1/16384 | 2187 | 1/2187 | 1/4782969 | 78125 |
| 8 | 1/256 | 1/65536 | 6561 | 1/6561 | 1/43046721 | 390625 |
| 9 | 1/512 | 1/262144 | 19683 | 1/19683 | 1/387420489 | 1953125 |
| … | … | … | … | … | … | … |

**Note**: Let $r_0=1$, the area of the initiator of the Sierpinski gasket is $A_0=\cos(\pi/3)\sin(\pi/3)\approx 0.433$; on the other, let $A_0=1$, the sidelength of the initiator is $r_0=[2/\sin(\pi/3)]^{1/2}\approx 1.5197$. However, both $r_0$ and $A_0$ can be taken as 1 for simplicity.

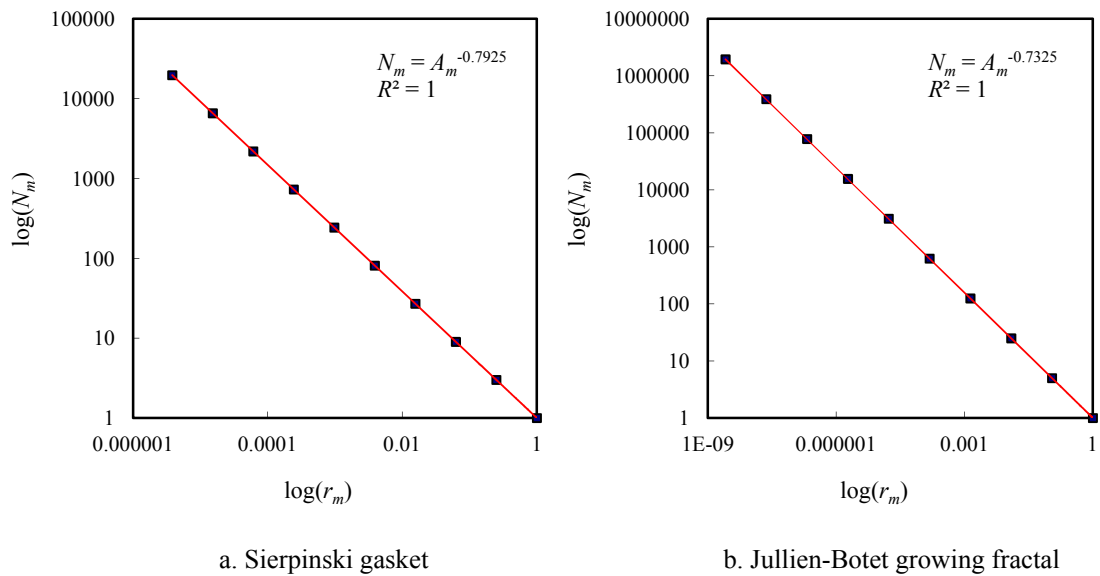

a. Sierpinski gasket        b. Jullien-Botet growing fractal

**Figure 3 The log-log plots of the scaling relations between area and number of fractal copies (the**

## 2.3 Fractal hierarchies of cities

Zipf's law of city-size distribution is equivalent to a hierarchical scaling law (Chen, 2012a). If a set of cities in a region are arranged into a hierarchy with cascade structure, which is illustrated in Figure 4, the hierarchical scaling relation can be expressed as follows

$$N_m = \eta S_m^{-1/q} = \eta S_m^{-p}, \qquad (7)$$

where $N_m$ refers to the number of cities in the $m$th level, and $S_m$ to the average size of the $N_m$ cities, $\eta = N_1 S_1^p$ is a proportionality coefficient, and the remaining notation is the same as in equations (1) and (2). Comparing equation (7) with equation (4) shows that the hierarchy of cities is similar to the fractal hierarchies. The corresponding relationships of measurements and parameters are tabulated as below (Table 2). This suggests that if we use urban area to represent city size, and the dimension of the urban area is $d=2$, then we will have $p=1/q=D_n/2$, where $D_n$ is the fractal dimension of network of cities corresponding to the hierarchy of cities.

Table 2 The corresponding relationships of measurements and parameters between regular fractal models and the model of hierarchy of cities

| Item | Fractal hierarchy | Hierarchy of cities |
| --- | --- | --- |
| Element | Fractal copies at the $m$th step | Cities at the $m$th level |
| Number | Number of fractal copies $N_m$ | Number of cities $N_m$ |
| Size measurement | Area of fractal copies $A_m$ | Average city size $S_m$ |
| Proportionality coefficient | $\mu = N_1 A_1^{D/2}$ | $\eta = N_1 S_1^p$ |
| Scaling exponent | $D/2$ | $p=1/q=D_n/D_s$ |

**Note**: It will be demonstrated that $D_n$ refers to the fractal dimension of a network of cities, and $D_s$ to the average value of the fractal dimension of size measurements of all cities in the network.

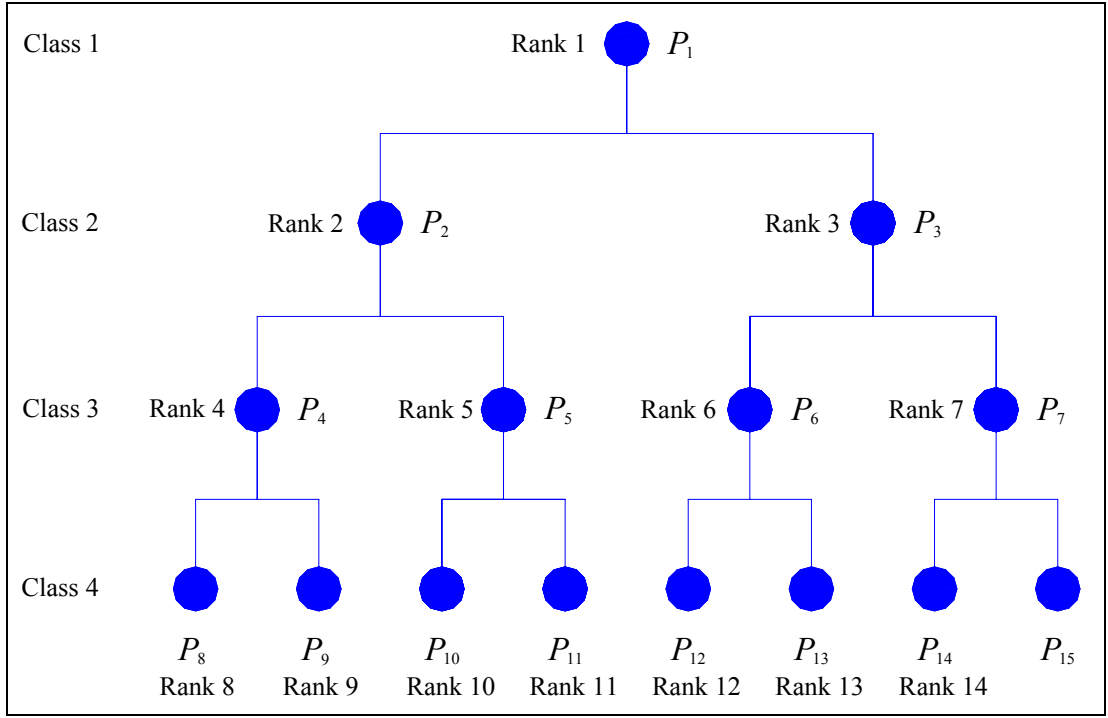

**Figure 4 A schematic diagram of hierarchy of cities with cascade structure (the first 4 levels)**

### 2.4 Spatial implication of Pareto scaling exponent

A fractal can be described with a power function, and a power function suggests a proportional relationship between two correlate measurements. If and only if the dimension of one measurement is identical to that of another measurement, the two measurements will be proportional to one another. This indicates the principle of dimensional homogeneity. Suppose that the dimension of the city number $N_m$ is $D_n$, and the dimension of city size $S_m$ is $D_s$. By the principle of dimensional consistency, a geometric measure relation can be constructed as

$$N_m^{1/D_n} \propto S_m^{-1/D_s}, \tag{8}$$

which bears an analogy to equation (5). Comparing equation (8) with equation (7) yields a parameter relation as below:

$$p = \frac{1}{q} = \frac{D_n}{D_s}, \tag{9}$$

which can be equivalently expressed as

$$q = \frac{1}{p} = \frac{D_s}{D_n}. \tag{10}$$

This suggests that the Pareto scaling exponent is the ratio of the dimension of city number to that of city size. Accordingly, the Zipf scaling exponent is the ratio of the dimension of city size to that of city number.

The dimension of the number of fractal copies is just the dimension of the fractal. Analogously, the dimension of city number $D_n$ is just the dimension of the network of cities. Therefore, the $D_n$ is a global parameter. The dimension of city size $D_s$ is concept of statistical average. In this sense, it is also a global parameter. However, the size dimension is related to the local measurements. In fact, each city's size corresponds to a dimension. The $D_s$ is the mean value of the dimensions of sizes of all cities. For the cities within a region, the $D_n$ value is determinate and invariant for a period of time. For each city, the dimension of city size is not determinate, but the average value of all the dimensions of city sizes is very stable. Thus the Pareto scaling exponent as well as the Zipf scaling exponent is resistant to change of values (Madden, 1956; Knox, 1994; Pumain, 1997).

Both Pareto distribution and Zipf's law indicate hierarchical scaling, which can be associated with spatial scaling. We can understand the spatial scaling through maps because a process of geographic mapping is just a scaling process. Suppose the area of a city within the urban boundary is $A$. The city can be represented with a circle of equal area. Thus we have

$$A = \pi R^2, \qquad (11)$$

where $R$ is the radius of the circle. The Pareto distribution can be equivalently expressed as

$$N(A) = CA^{-\alpha} = \frac{C}{\pi^{\alpha}} R^{-2\alpha} = N(R), \qquad (12)$$

where $N(A)$ refers to the number of the cities with area greater than or equal to $A$, and $N(R)$ to the number of the cities with radius greater than or equal to $R$, $C$ is a proportionality constant, and $\alpha$ is the scaling exponent measured with urban area. According to the geometrical measure relation, we have

$$\alpha = \frac{D_n}{D_a} = \frac{D_n}{2}, \qquad (13)$$

in which the dimension $D_n$ corresponds to number $N(A)$, while the Euclidean dimension $D_a=2$ corresponds to the area measurement $A$. The cities can be displayed on a digital map. Suppose that the scale of the map is defined as below

$$s = \frac{L}{R}, \tag{14}$$

in which $s$ refers to the map scale, and $L$ to the radius of the city on the map. If a city's radius is greater than or equal to $R$, it will be shown on the map, or else it will be neglected. Then we will shown $N(R)$ cities on the map. Substituting equation (14) into equation (12) yields

$$N(s) = \frac{C}{\pi^\alpha}(\frac{L}{s})^{-2\alpha} \propto s^{2\alpha} = s^{D_n}. \tag{15}$$

This implies that the larger the map scale is, the more number of cities will be shown. The fractal dimension of a network of cities can be measured with the scale of map. The scaling relation based on the map scale is as follows

$$N(\lambda s) \propto (\lambda s)^{D_n} \propto \lambda^{D_n} N(s). \tag{16}$$

in which $\lambda$ denotes a scale factor. This suggests that if the scale of map varies from $s$ to $\lambda s$, the number of the cities which will appears on the map will change from $N(s)$ into $(\lambda^{D_n})N(s)$.

The urban population is often employed to measure city size. Empirically, the relation between city population and urban area takes on an allometric scaling (Batty and Longley, 1994; Chen, 2010; Lee, 1989). The allometric growth law can be expressed as

$$A = aP^b = aP^{2/D_p}. \tag{17}$$

where $a$ is a proportionality coefficient, $b=2/D_p$ is a scaling exponent. Equation (17) is theoretically equivalent to the following relation (Lee, 1989)

$$R \propto P^{b/2} \propto P^{1/D_p}. \tag{18}$$

Thus, in terms of equation (12), we have

$$N(P) = \frac{C}{\pi^\alpha}(aP^{1/D_p})^{-2\alpha} \propto P^{-D_n/D_p} = P^{-\gamma}. \tag{19}$$

Accordingly, the Pareto scaling exponent is

$$\gamma = \frac{D_n}{D_p}, \tag{20}$$

where $\gamma$ refers to the Pareto scaling exponent measured with city population size. This Pareto exponent used to be treated as the fractal dimension of city-size distribution.

## 3. Empirical evidences

### 3.1 Hierarchical scaling law

The hierarchical scaling law of cities will be employed to make two empirical analyses. The rank-size scaling is equivalent to the hierarchical scaling (Chen, 2012a), thus Pareto's law as well as Zipf's law can be replaced by the hierarchical scaling law such as

$$N(m) = CP(m)^{-D} = CP(m)^{-D_n/D_p}, \tag{21}$$

where $m$ represents the level order in a hierarchy, $N(m)$ refers to the city number in the $m$th level, and $P(m)$ to the average population size of the $N(m)$ cities. As for the parameters, $C=N(1)P(1)^D$ denotes the proportionality coefficient, and $D=D_n/D_p$ is the fractal dimension of city size distribution, i.e., the Pareto exponent. The rest symbols have be narrated above. Equation (21) is a theoretical expression based on continuous variable and population. For an observed dataset (sample) of cities, the discrete format of equation (21) is

$$N_m = CP_m^{-D} = CP_m^{-D_n/D_p}, \tag{22}$$

in which $N_m$ is the city number in the $m$th level, $P_m$ denotes the average population size of the $N_m$ cities, and the other notation fulfils the same role as in equation (21). In fact, equation (22) is a special case of equation (8). Substituting the general size measure $S$ with city population $P$, equation (8) will becomes equation (22).

Compared with Zipf's law, the hierarchical scaling law has two advantages. First, the similarity and difference between the hierarchical scaling and fractal scaling are clear. The hierarchical scaling is based on the cascade structure (Figure 4), which bears a clear analogy with the fractal structure (Figure 2). Both the regular fractal and hierarchy of cities have the same scaling processes and self-similar patterns. Second, the hierarchical series of cities has larger power against the disturbance of random noises of observed data than the rank-size series. The processing based on statistical average has a function of "filter" and denoising. Pareto's law gains the second advantage of Zipf's law, while the hierarchical scaling law wins the first advantage over Pareto's law.

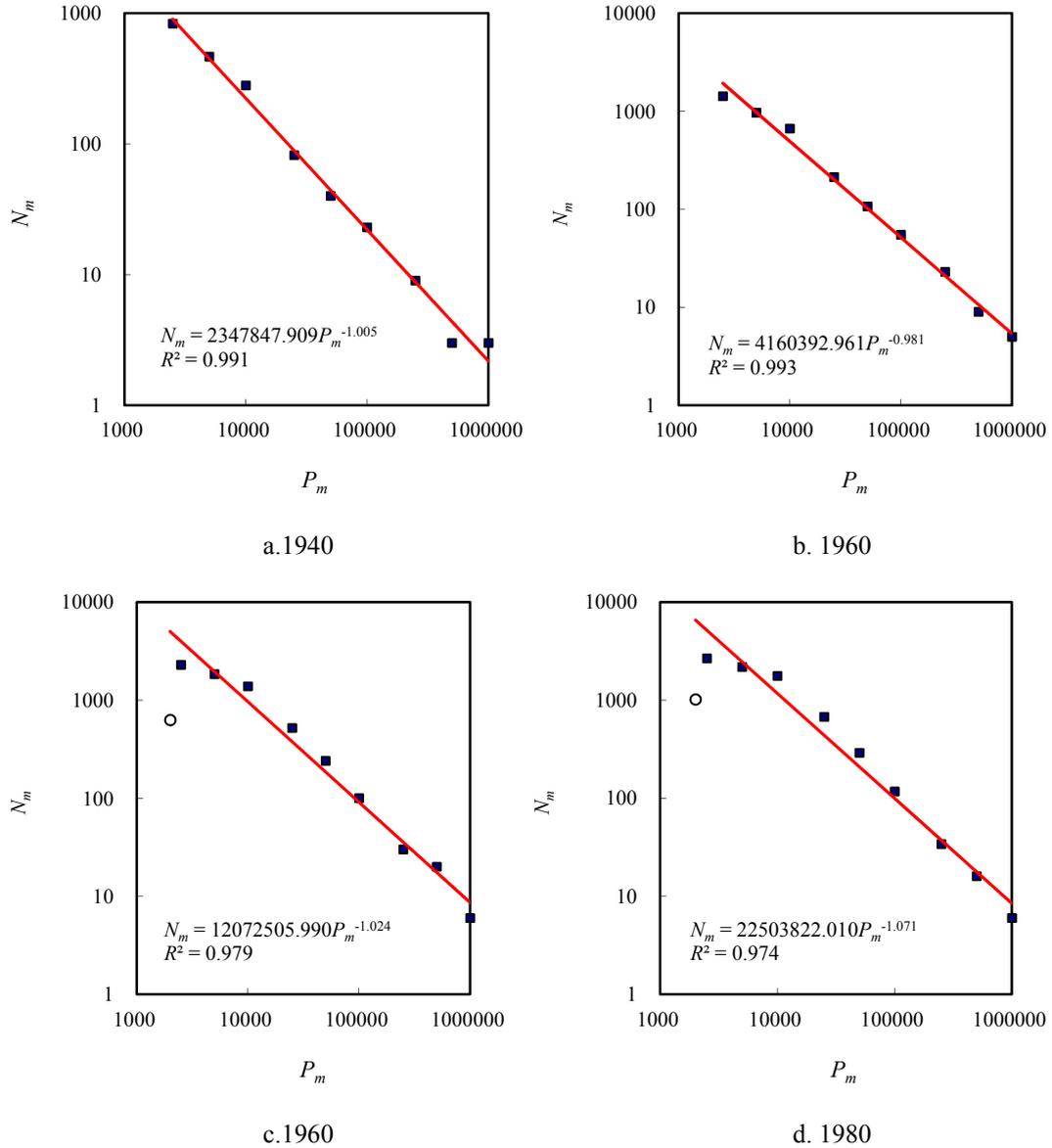

**Figure 5 The scaling relations between the population size and number of cities in the United States, 1900-1980 (by Chen, 2011)**

(**Note**: In 1960 and 1980, the last class is treated as an outlier because of undergrowth of cities. The circles indicate the outliers beyond the scaling ranges)

## 3.2 Cases of the U.S. cities

The first case is the hierarchy of cities in the United States. The data for the period of 1900 to 1980 were processed by King (1984) (Table 2). The relationships between the city number and population size follow the scaling law on the whole, and the fractal dimension values of city-size distributions, $\gamma$, have been estimated by Chen (2011) (Figure 5, Table 3). From 1900 to 1980, the Pareto scaling exponent $\gamma=D_n/D_p$ varied around 1. This suggests that the fractal dimension of networks of cities is very close to the mean dimension of the city population. The dimension of city population is unknown; therefore, the fractal dimension of the network of cities in U.S. cannot be evaluated. However, this example is helpful for our understanding the spatial development of the urban system. If the average dimension value of city population is $D_p=2$, the fractal dimension values of the network of the U.S. cities are about $D_n=2.0106$ (1900), $D_n=1.9626$ (1940), $D_n=2.0485$ (1960), and $D_n=2.1416$ (1980), respectively; If the dimension of city population is $D_p=1.7$, the fractal dimensions of the network of cities are about $D_n=1.7090$ (1900), $D_n=1.6682$ (1940), $D_n=1.7412$ (1960), and $D_n=1.8204$ (1980). The increase of the fractal dimension of the urban systems indicates the processes of birth of new cities and growth of space-filling extent.

**Table 3 The scaling exponents, the corresponding goodness of fit, and the estimated fractal dimension for the US cities, 1900-1980**

| Year | Scaling exponent $\gamma$ | Goodness of fit $R^2$ | $D_p=2$ | $D_p=1.7$ |
|---|---|---|---|---|
| 1900 | 1.0053 | 0.9909 | 2.0106 | 1.7090 |
| 1940 | 0.9813 | 0.9931 | 1.9626 | 1.6682 |
| 1960 | 1.0243 | 0.9786 | 2.0485 | 1.7412 |
| 1980 | 1.0708 | 0.9736 | 2.1416 | 1.8204 |

### 3.3 Cases of Indian cities

The same scaling analysis can be applied to Indian cities, which satisfy Zipf's distribution (Gangopadhyay and Basu, 2009). This is the second case of this paper. The census data of Indian cities during the period of 1981 to 2001 have been processed by Basu and Bandyapadhyay (2009). The fractal dimension values of Indian city-size distribution have been estimated by Chen (2012) (Figure 6; Table 4). The same problems arise that we know nothing about the dimension of city population. If the average dimension value of city population is $D_p=2$, the fractal dimension values of the network of Indian cities are about $D_n=2.2945$ (1981), $D_n=2.2174$ (1991), and $D_n=2.1055$ (2001); If the dimension of city population is $D_p=1.7$, the fractal dimensions of the network of cities are around $D_n=1.9503$ (1981), $D_n=1.8848$ (1991), and $D_n=1.7897$ (2001), respectively. The decrease of the fractal dimension of the urban systems indicates that the dimension of Indian city population cannot be assumed to be fixed. The dynamics of Indian urban evolution is different from that of U.S. The common character is that the fractal dimension values of networks of cities approaches to the average dimension values of city population.

**Table 4 The scaling exponents, the corresponding goodness of fit, and the estimated fractal dimension for Indian cities, 1981-2001**

| Year | Scaling exponent $\gamma$ | Goodness of fit $R^2$ | $D_p=2$ | $D_p=1.7$ |
| --- | --- | --- | --- | --- |
| 1981 | 1.1472 | 0.9982 | 2.2945 | 1.9503 |
| 1991 | 1.1087 | 0.9967 | 2.2174 | 1.8848 |
| 2001 | 1.0528 | 0.9900 | 2.1055 | 1.7897 |

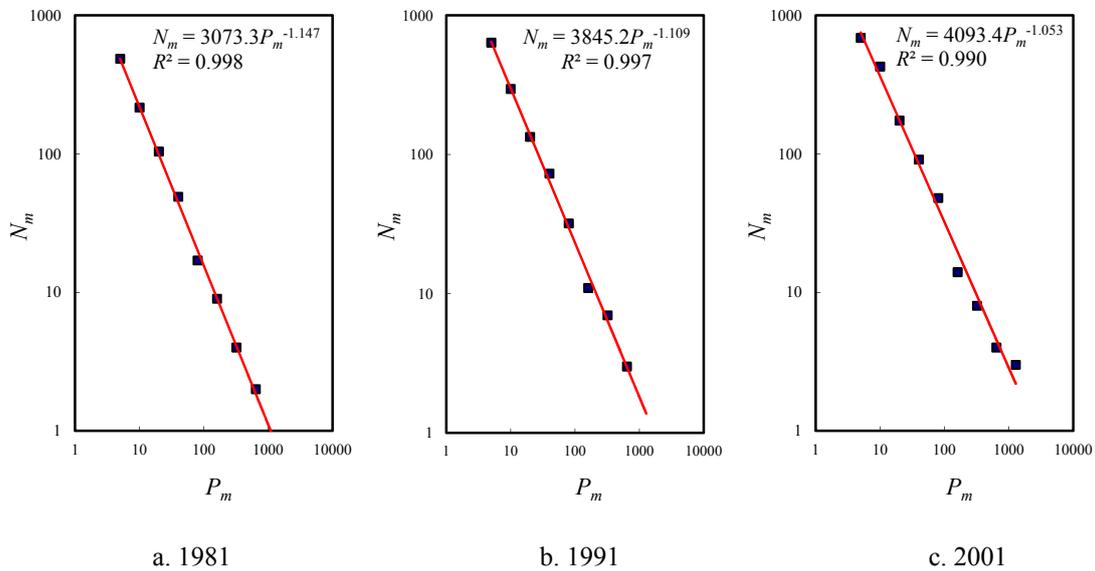

a. 1981  b. 1991  c. 2001

**Figure 6 The scaling relations between the lower limit of population size and the number of cities in India, 1981-2001 (by Chen, 2012)**

## 4. Questions and discussion

Using the theory developed in this paper, we can answer at least three questions. The first is how to understand the fractal property of city-size distributions. The rank-size distribution of cities is equivalent to an urban hierarchy with cascade structure, which is in turn equivalent to a self-similar network. Pareto's law is mathematically equivalent to a hierarchical scaling law, which can be described with an inverse allometric relation. The second is how to understand the scaling exponent of a city-size distribution. Because the size measure is not a linear scale, the Pareto exponent is not the real fractal dimension, but a ratio of one fractal dimension to the other. The third is how to understand the long-term stability of the Zipf's distribution. The Zipf exponent equals the reciprocal of Pareto exponent, which is the ratio of the fractal dimension of network of cities to the average dimension of city size. The dimension of a city network is stable for a long time. The dimension value of a city is not stable, but the average value of the dimensions of all the cities in a network approaches a constant. Thus the Pareto exponent as well as the Zipf exponent has no significant change from year to year.

Now, a new question arises, that is how to comprehend the property of the dimensions of city network and size. Suppose that the city size is measured with urban population. If the spatial distribution of a city's population follows Clark's law (Clark, 1951), the dimension of urban population is $D_p=d=2$ (Chen and Feng, 2012); if the urban population distribution follows Smeed's law (Smeed, 1963), the dimension of city population is a fractional value ranging from 0 to 2 (Batty and Longley, 1994). In many cases, the spatial distributions of urban population follow Clark's law, thus the fractal dimension of city population is $D_p=2$. If so, it will be hard to understand the fractal dimension values of the U.S. networks of cities. Especially, the dimension values of Indian network of cities cannot be explained using the conventional concept of geographical space. If we define a network of cities in a 2-dimensional space based on a digital map, the fractal dimension of the network must come between 0 and 2. However, many fractal dimensions of city networks go beyond the upper limit if the urban population is a 2-dimensional measure of size (i.e., $D_p=2$).

In this instance, the notion of generalized space should be introduced into geography. The urban area-population allometric growth is a simple and good example to illustrate the types of geographical space. Using the allometric scaling relation between urban area and size, we can derive three concepts of geographical space. Based on the size measure of urban population, equation (17) can be replaced by

$$A = aP^b = aP^{D_a/D_p}, \qquad (23)$$

in which $a$ refers to the proportionality coefficient, and $b=D_a/D_p$ to the scaling exponent. Given different spatio-temporal conditions, equation (23) will result in three types of geographical space

(Table 5).

**Table 5 Three types of geographical space: real space, phase space, and order space**

| Space | Description | Physical base and data | Basic fractal dimension | Dimension value range |
|---|---|---|---|---|
| Real space (R-space) | Empirical space | Spatial series or random observational data based on maps, digital maps, remotely sensed images, etc. | Box dimension, radial dimension | $0 \leq D \leq 2$ |
| Phase space (P-space) | Abstract space | Temporal series based on daily/monthly/yearly observations and measurements, etc. | Similarity dimension, correlation dimension | $0 \leq D \leq 3$ |
| Order space (O-space) | Abstract space | Cross-sectional data based on regional observations and measurement, etc. | Similarity dimension | $0 \leq D \leq 3$ |

The first is the real space (R-space). For a given city at certain time, equation (23) should be substituted by

$$A(r) = a'P(r)^{b'} = a'P(r)^{D'_a / D'_p}, \qquad (24)$$

where $r$ denotes the radius from the city center, $A(r)$ refers to the land-use area within a radius of $r$ unit from the center ($0 \leq r \leq R$, where $R$ is the maximum radius of a cities), and $P(r)$ to the population within the same sphere as $A(r)$, $a'$ is the proportionality coefficient, and $b'=D_a'/D_p'$ is the scaling exponent. Equation (24) can be derived from two fractal models as follows

$$A(r) = A_0 r^{D'_a}, \qquad (25)$$

$$P(r) = P_0 r^{D'_p}, \qquad (26)$$

where $A_0$ and $P_0$ are two proportionality constants, $D_a'$ is the fractal dimension of urban land use form, and $D_p'$ is the dimension of population distribution of the city. This suggests that the fractal dimensions $D_a'$ and $D_p'$ belong to the real geographical space ($0 \leq D_a', D_p' \leq 2$).

The second is the phase space (P-space). For a given city within a period of $n$ years, equation (23) should be replaced with

$$A(t) = a''P(t)^{b''} = a''P(t)^{D''_a / D''_p}, \qquad (27)$$

where $t$ denotes the year ($t$=1, 2, …, $n$), $A(t)$ refers to the land-use area in the $t$th year within a radius of $R$ unit from the center, and $P(t)$ to the population in the same year within the same sphere as $A(t)$, $a''$ is the proportionality constant, and $b''=D_a''/D_p''$ is the scaling exponent. Equation (27) can be derived from two models such as

$$A(t) = A_T R(t)^{D_a''}, \tag{28}$$

$$P(t) = P_T R(t)^{D_p''}, \tag{29}$$

where $A_T$ and $P_T$ are two proportionality constants, $R(t)$ is the largest radius of a city in the $t$th year, $D_a''$ is the average fractal dimension of urban land use form in the $n$ year, and $D_p''$ is the average dimension of population distribution of the city in the same period. If the area within an urban boundary is $A$, the largest radius can be defined by $R=(A/\pi)^{1/2}$. This implies that the fractal dimensions $D_a''$ and $D_p''$ belong to a generalized geographical space—phase space ($0 \leq D_a'', D_p'' \leq 3$).

The third is the order space (O-space). For $N$ cities within a region in given year, equation (23) should be replaced by

$$A(k) = a''' P(k)^{b'''} = a''' P(k)^{D_a'''/D_p'''}, \tag{30}$$

where $k$ denotes the rank of a city ($k$=1, 2, …, $N$), $A(k)$ refers to the land-use area within an urban boundary, and $P(k)$ to the population inside the same urban boundary, $a'''$ is the proportionality coefficient, and $b'''=D_a'''/D_p'''$ is the scaling exponent. Equation (30) can be derived from two Zipf's laws

$$A(k) = A_1 k^{-D_a'''/D_n}, \tag{31}$$

$$P(k) = P_1 k^{-D_p'''/D_n}, \tag{32}$$

where $A_1$ and $P_1$ are two proportionality constants, $D_a'''$ is the average fractal dimension of urban land use form of the $N$ cities, and $D_p'''$ is the average dimension of population distribution of the same urban system. This implies that the fractal dimensions $D_a'''$ and $D_p'''$ belong to another generalized geographical space—order space($0 \leq D_a''', D_p''' \leq 3$). Equation (30) can equivalently expressed as the following hierarchical scaling relation

$$A(m) = a''' P(m)^{b'''} = a''' P(m)^{D_a'''/D_p'''}. \tag{33}$$

In theory, the same kind of fractal dimension of different spaces should be equal to one another. For a given city at a given time ($t$ is determined), if the urban radius is defined according to certain criterion ($r=R$), we have

$$b = \frac{D'_a}{D'_p} = \frac{D''_a}{D''_p} = \frac{D'''_a}{D'''_p}. \tag{34}$$

However, because of random disturbance and varied human factors, the observed data do not always support this equation. In practice, an approximate relation is as below:

$$b \approx \frac{D'_a}{D'_p} \approx \frac{D''_a}{D''_p} \approx \frac{D'''_a}{D'''_p}. \tag{35}$$

This relation can be testified with the statistical average of large-sized samples.

For the Pareto distribution, the fractal dimension of city network $D_n$ and the average dimension of city population $D_p$ belong to the order space rather than the real space. The value of these dimension come between 0 and 3 instead of varying from 0 to 2. The reason is that urban form, population, networks of cities, and so on, are all defined in a 3-dimensional space. However, the real space is defined in a 2-dimensional space based on digital maps. The well-known regular fractals, Sierpinski gasket (Figure 1) and the Jullien-Botet growing fractal (Figure 2), can be employed to illustrate this dimension difference between the real space and order space. The relation between the area $A_m$ and the length of its external/interior boundary $L_m$ of the two fractals is an inverse allometric scaling (Chen, 2010). The fractal measure relation can be expressed as

$$A_m = \xi L_m^{-v} = \xi L_m^{-D_a/D_l}, \tag{36}$$

where the parameters $\xi = A_1 L_1^v$, $v = D_a/D_l = \ln(A_m/A_{m-1})/\ln(L_{m-1}/L_m)$, $D_a$ is the fractal dimension of the fractal form, and $D_l$ is the fractal dimension of the external/interior boundary.

The Jullien-Botet growing fractal is very simple and clear. In the real space, the box dimensions of this growing fractal form and its external boundary are $D_a = D_l = \ln(5)/\ln(3) \approx 1.465$. Accordingly, the scaling exponent $v = D_a/D_l = 1$. On the other, in the order space, the similarity dimensions of the Jullien-Botet growing fractal form and its boundary are also $D_a^* = D_l^* = \ln(5)/\ln(3) \approx 1.465$. Correspondingly, the scaling exponent $v^* = D_a^*/D_l^* = 1$. For this fractal, the ratio of one dimension to the other based on the real space equals that based on the generalized space, i.e., $v = v^*$.

However, for the Sierpinski gasket, the case is different and complicated to some extent. In the real space, the fractal dimensions of the fractal form and the interior boundary can be determined with box-counting method, and we have

$$D_a = \frac{\ln N_a(\varepsilon_m)}{\ln \varepsilon_m} = \frac{\ln(3)}{\ln(2)} \approx 1.585, \quad D_l = -\lim_{m \to \infty} \frac{\ln N_l(\varepsilon_m)}{\ln \varepsilon_m} = \frac{\ln(3)}{\ln(2)} \approx 1.585,$$

where $N_a(\varepsilon_m)$ is the least number of nonempty boxes for the fractal object, $N_l(\varepsilon_m)$ is the least number of nonempty boxes for the interior boundary, and $\varepsilon_m$ is the linear scale of the boxes. Thus the allometric scaling exponent is $v=D_a/D_l=1$. However, in the order space, the box dimensions of the fractal form and the interior boundary should be replaced by the corresponding similarity dimensions

$$D_a^* = D_a = -\frac{\ln(N_m/N_{m-1})}{\ln(\varepsilon_m/\varepsilon_{m-1})} = \frac{\ln(3)}{\ln(2)} \approx 1.585, \quad D_l^* = -\frac{\ln(N_m^*/N_{m-1}^*)}{\ln(\varepsilon_m/\varepsilon_{m-1})} = \frac{\ln(5)}{\ln(2)} \approx 2.3219,$$

where $N_m$ is the number of fractal copies of the gasket, $N_m^*$ is the number of fractal copies of its interior boundary, and $\varepsilon_m$ is the linear scale of the fractal copies. Obviously, the fractal dimension of the fractal line exceeds the upper limit of the Euclidean dimension of the embedding space (see Appendix part). Thus the allometric scaling exponent is $v^*=D_a^*/D_l^* \approx 1.585/2.3219 \approx 0.6826 < 1 = v$. This indicates that the scaling exponent based on the real space is not always equal to that based on the generalized space.

The generalized space can be utilized to explain the difference between the urban evolution of the U.S. cities and that of Indian cities. If we fix the average dimension of the city population, the fractal dimension of network of the U.S. cities went up from 1900 to 1980 as a whole. This is easy to understand since the fractal dimension is a measure of space-filling. This suggests the average value of the fractal dimension of city population is stable, and in the mass, the city number became larger and larger. As for Indian cities, the case is different. The fractal dimension of the network of cities went up since the city number increased. However, the average dimension of city population increased faster than the fractal dimension of city network. In other words, Indian urban population density went faster than the density of the spatial distribution of cities. An inference is that more and more high-rise buildings appeared in Indian cities to accommodate more and more urban inhabitants so that the fractal dimension of city population went up and up from 1981 to 2001.

## 5. Conclusions

The rank-size rule is a very simple scaling law followed by many observations of the ubiquitous empirical patterns in physical and social systems. The rank-size distribution can be formulated with Zipf's law or Pareto's law. Krugman (1996) once said, "The usual complaint about economic theory is that our models are oversimplified -- that they offer excessively neat views of complex, messy reality. (In the case of the rank-size distribution) the reverse is true: we have complex, messy models, yet reality is startlingly neat and simple." Now, we can see that reality is simple, but idea is profound. Based on the mathematical derivation, empirical analysis, and theoretical generalization, the main conclusions of this paper can be drawn as follows.

**First, the Pareto scaling exponent of a rank-size distribution of cities is a ratio of the fractal dimension of a network of cities to the average dimension of city population within the network.** Accordingly, the Zipf scaling exponent is the ratio of the mean dimension of city population to the fractal dimension of city network. The rank-size distribution of cities is equivalent to a self-similar hierarchy of cities, and a hierarchy with cascade structure is equivalent to a network with fractal structure. The fractal network of cities can be described with a fractal dimension. On the other land, within the network, the population distribution of each city has a dimension. Different cities have different dimension values of urban population, but the average value of the population dimensions approaches to a constant. The fractal dimension of city network is very stable. Thus the Pareto/Zipf exponent is steadfast for a long time.

**Second, the fractal dimension of a network of cities and the dimension of city population based on the Pareto distribution are defined in a generalized space rather real space.** The rank-size distribution of cities can be used to define a new geographical space, which is a kind of generalized space. The Pareto exponent as well as the Zipf exponent is a ratio of two dimensions defined in the generalized space. The fractal dimension of a network of cities defined in a real space can be determined with the box-counting method, and the dimension value of urban population of each city can be estimated with the mass-radius scaling or the box-counting method. In theory, the fractal dimension of real space equals the corresponding fractal dimension of the generalized space. However, in empirical studies, the fractal dimension of a real space approximates to the dimension of the corresponding generalized space but there always are a few

errors. In the sense of statistical average, the dimension of a generalized space can be testified with the observed values of the fractal dimension of the corresponding real space.

**Acknowledgment**

This research was sponsored by the National Natural Science Foundation of China (Grant No. 41171129). The supports are gratefully acknowledged.

# Appendix

The Sierpinski gasket is a self-similar hierarchy (Figure A). The linear scale sequence is 1, 1/2, $1/2^2$, …, $1/2^{m-1}$ ($m$=1,2,3,…), and the corresponding number sequence of fractal copies or nonempty boxes is 1, 3, $3^2$, …, $3^{m-1}$. The nonempty box number equals the fractal copy number. Thus the fractal dimension of the gasket is $D_a = D_a^* = \ln(3)/\ln(2) \approx 1.585$. However, for the interior boundary, the nonempty box number does not equal the fractal copy number (Table A). The number sequence of fractal copies is 1, 5, $5^2$, ... , $5^{m-1}$, while the number sequence of fractal nonempty boxes is 1, 5, 19, ... , $3^m$. Thus the similarity dimension of the fractal curve is $D_l^*=\ln(5)/\ln(2)\approx 2.3219$, and if $m\rightarrow\infty$, the box dimension will be $D_l^*=\ln(3^m)/\ln(2^{m-1})=[m/(m-1)]*\ln(3)/\ln(2)\rightarrow 1.585$.

**Table A The order, linear scale of fractal copies, number of fractal copies, and the number of boxes covering the interior boundary**

| Order $m$ | Linear scale $\varepsilon_m$ | Fractal form | | Interior boundary | |
|---|---|---|---|---|---|
| | | Number of fractal copies $N_{am}$ | Box number $N_a(\varepsilon_m)$ | Number of fractal copies $N_{lm}$ | Box number $N_l(\varepsilon_m)$ |
| 1 | 1 | 1 | 1 | 1 | 1 |
| 2 | 1/2 | 3 | 3 | 5 | 5 |
| 3 | 1/4 | 9 | 9 | 25 | 19 |
| 4 | 1/8 | 27 | 27 | 125 | 65 |
| 5 | 1/16 | 81 | 81 | 625 | 211 |
| 6 | 1/32 | 243 | 243 | 3125 | 665 |
| 7 | 1/64 | 729 | 729 | 15625 | 2059 |
| 8 | 1/128 | 2187 | 2187 | 78125 | 6305 |
| 9 | 1/256 | 6561 | 6561 | 390625 | 19171 |
| 10 | 1/512 | 19683 | 19683 | 1953125 | 58025 |
| 11 | 1/1024 | 59049 | 59049 | 9765625 | 175099 |
| 12 | 1/2048 | 177147 | 177147 | 48828125 | 527345 |
| 13 | 1/4096 | 531441 | 531441 | 244140625 | 1586131 |
| 14 | 1/8192 | 1594323 | 1594323 | 1220703125 | 4766585 |
| 15 | 1/16384 | 4782969 | 4782969 | 6103515625 | 14316139 |
| ... | ... | ... | ... | ... | ... |

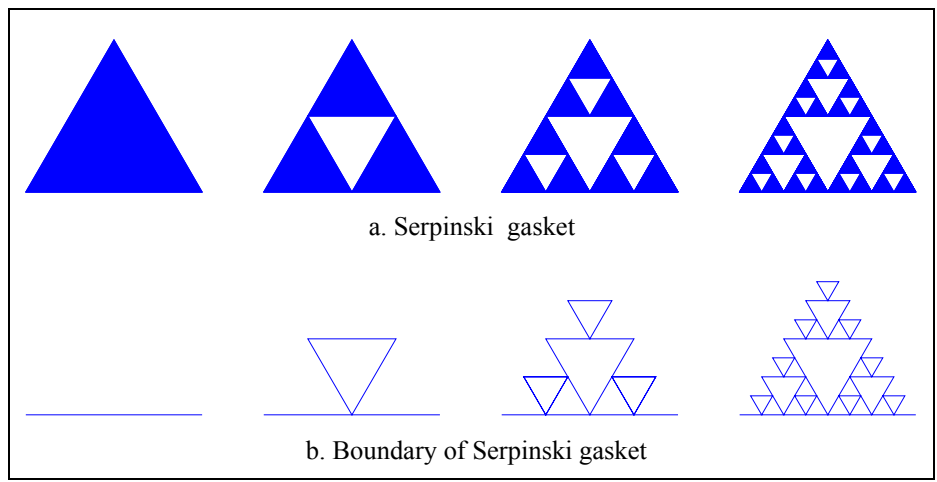

a. Serpinski gasket

b. Boundary of Serpinski gasket

**Figure A The Sierpinski gasket and its interior boundary curve (the first four steps)**